\documentclass{article}
\usepackage[letterpaper, scale=0.8]{geometry}
\usepackage{amsmath}
\usepackage{amsfonts}
\usepackage{physics}
\usepackage{graphicx}
\usepackage{subcaption}
\usepackage{float}
\usepackage{booktabs}
\usepackage{multirow}
\usepackage{algorithm}
\usepackage{algorithmic}
\usepackage{cite}
\usepackage{authblk}

\newcommand{\keywords}[1]{%
  \begin{flushleft}
    \textbf{Keywords:} #1
  \end{flushleft}
}

\title{Operator Learning for Reconstructing Flow Fields from Sparse Measurements: an Energy Transformer Approach}

\author[1]{Qian Zhang}
\author[2]{Dmitry Krotov}
\author[1]{George Em Karniadakis\thanks{george\_karniadakis@brown.edu}}
\affil[1]{Division of Applied Mathematics, Brown University}
\affil[2]{MIT-IBM Watson AI Lab and IBM Research}
\date{}

\begin{document}
\maketitle

\begin{abstract}
    Machine learning methods have shown great success in various scientific areas, including fluid mechanics. However, reconstruction problems, where full velocity fields must be recovered from partial observations, remain challenging. In this paper, we propose a novel operator learning framework for solving reconstruction problems by using the Energy Transformer (ET), an architecture inspired by associative memory models. We formulate reconstruction as a mapping from incomplete observed data to full reconstructed fields. The method is validated on three fluid mechanics examples using diverse types of data: (1) unsteady 2D vortex street in flow past a cylinder using simulation data; (2) high-speed under-expanded impinging supersonic jets impingement using Schlieren imaging; and (3) 3D turbulent jet flow using particle tracking. The results demonstrate the ability of ET to accurately reconstruct complex flow fields from highly incomplete data (90\% missing), even for noisy experimental measurements, with fast training and inference on a single GPU. This work provides a promising new direction for tackling reconstruction problems in fluid mechanics and other areas in mechanics, geophysics, weather prediction, and beyond.
\end{abstract}



\section{Introduction}

Machine learning methods have been widely applied in various scientific problems, e.g., physics~\cite{price2024weather,AlphaTensor2022, aad2012observation,cai2022review, carleo2017solving,jin2020SympNets,zhang2024sms,zhang2023artificial,theilman2024loihi,wu2023comprehensive,wan2024rfg}, chemistry~\cite{jumper2021highly,merchant2023scaling, cao2023lno}, biology~\cite{zhang2022sysbio, zhang2022AOSLO}, epidemiology~\cite{zhang2021PLOS, kharazmi2021}, etc. This great success is due to the rapid simultaneous development of algorithms and computation power, which enable transforming the tasks that are difficult or expensive to solve by traditional methods into optimization tasks, which may be easier to tackle.

In this paradigm, operator learning is a prominent approach that has gained increasing interest recently~\cite{lu2021learning, cao2023lno, goswami2022physics, kahana2022spiking,zhang2022hints}. In traditional methods, time-consuming and computationally expensive algorithms need to be run every time for a new input condition. On the other hand, operator learning aims to learn the mapping from general input conditions or distributions to solutions, i.e. the solution operator, by a neural network. Hence, neural operators can be used as surrogates to the traditional methods as the forward pass of a neural network usually is much faster than the conventional algorithms. Some recent progress in architectures like DeepONet~\cite{lu2021learning}, Fourier neural operator (FNO)~\cite{li2020fourier}, LNO~\cite{cao2023lno} and ViTO~\cite{ovadia2024vito} have demonstrated the great potential of operator learning in diverse applications. More specifically, DeepONet is a mesh-free method, allowing solutions to be evaluated at any point in the domain. FNO leverages the Fourier transform to efficiently model long-range dependencies and capture global structures in data. LNO is based on Laplace transform and excels in handling transient responses and non-periodic signals. ViTO utilizes the Vision Transformer to learn the operator from data on a regular mesh, offering a fast and efficient approach.

However, the situation is more challenging for reconstruction problems, i.e. full flow fields, from sparse and sporadic measurements. A common example is jet flow reconstruction, where measurement limitations often result in only partial access to the flow field. Although obtaining high-resolution flow fields is desirable, it can be costly. Therefore, the primary goal in these scenarios is to reconstruct the complete data from the observed, incomplete data. In addition, the objective of the operator learning method for reconstruction problems is to learn the mapping between the observed data and the full data with limited example pairs. Clearly, the incomplete and noisy nature of the data pose great challenges for the generalization ability of operator learning methods, because the test data is probably out of the distribution of the training data. For instance, the physics-informed diffusion model (PIDM)~\cite{shu2023pidm} effectively reconstructs flow fields from sparse measurements. However, this approach relies on measurements taken at regular mesh points, which limits its applicability to more general and irregular scenarios. Expanding operator learning methods to handle such cases remains an open area of research. For non-operator learning methods, Physics-Informed Neural Networks (PINNs)~\cite{raissi2019physics} have shown some success in reconstruction problems in fluid mechanics~\cite{jin2021nsfnets, raissi2020hidden, cai2022review, cai2021aiv, cai2024ptv, boster2023aiv, toscano2024inferring}, because PINNs can incorporate the prior knowledge (physical laws) into the loss function. However, they are still limited by the requirement of the explicit form of the governing equations, relatively high computational cost and lack of ability to generalize.

Herein, we propose to use the Energy Transformer \cite{hoover2024energy} for solving reconstruction problems in the operator learning framework. The Energy Transformer is inspired by recent developments in Dense Associative Memory models \cite{krotov2016dense,krotov2023new}, which are high memory storage capacity extensions of Hopfield networks \cite{hopfield1982neural}. Energy Transformer reconstructs the patterns from partial data and fill in the missing parts, which concurs with the objective of reconstruction problems. This makes Energy Transformer an ideal model for reconstruction operator. More specifically, the patterns are stored as the local minima of the energy function learned by the Energy Transformer. Then in reconstruction, the full data are inferred by minimizing the energy function starting from the observed data.

In this paper, we formulate the reconstruction problems in an operator learning framework with the Energy Transformer. We present three diverse examples in fluid mechanics using multimodal data, including real problems with noisy data, to demonstrate the effectiveness of the method.

\section{Method}
\subsection{Operator learning framework}
In this section, we formulate an operator learning framework for the reconstruction problem and introduce the proper notation.
We denote a data sample as a collection of position-value pairs $\mathcal{P}_a=\{(x_a^A, v_a^A)\}_{A=1}^{M_a}$, where $x_a^A$ represents the position, $v_a^A$ denotes the data value at position $x_a^A$, and $M_a$ is the number of position-value pairs in sample $a$. This collection $\mathcal{P}_a$ is referred to as a \textbf{full} data sample. Meanwhile, an \textbf{observed} data sample is defined as $\tilde{\mathcal{P}}_a=\{(x_a^A, v_a^A)\}_{j\in\mathcal{O}_a}$, where $\mathcal{O}_a$ indexes the observed position-value pairs. A \textbf{dataset} is defined as a collection of full data samples $\mathcal{D}=\{\mathcal{P}_a\}_{a=1}^{M}$, where $M$ is the number of samples. The goal of operator learning for the reconstruction problem is to learn a mapping $\mathcal{R}$ from the observed data sample to the full data sample, i.e.,
\begin{equation}
    v_{a'}^A = \mathcal{R}(x_{a'}^A; \tilde{\mathcal{P}}_a), \quad \forall a'\not\in\mathcal{O}_a.
\end{equation}

In certain reconstruction problems, we have access to adequate full data samples, making a data-driven approach possible. This involves representing the reconstruction operator with a neural network $\mathcal{R}_\theta$. The observed data for training can be easily obtained by selecting some pairs from the full data samples. Furthermore, for new observed data, the neural network can predict the full data directly without running complex reconstruction algorithms, which is the advantage of operator learning compared to PINNs reconstruction or any other method \cite{cai2021aiv}.

We assume that the full data sample can be divided into patches of the same size, with the observed data consisting of some patches from the full data (i.e., the remaining patches are masked). Therefore, the positions $x_a^A$ become the patch indices. The model is expected to learn to fill in the masked patches based on the observed data. The key idea is to use the Energy Transformer (introduced in detail in the next section) to serve as a memory model to store the patterns of the full data, and retrieve the missing parts by querying the memory with the observed data.

\begin{figure}[H]
    \centering
    \includegraphics[width=0.8\textwidth]{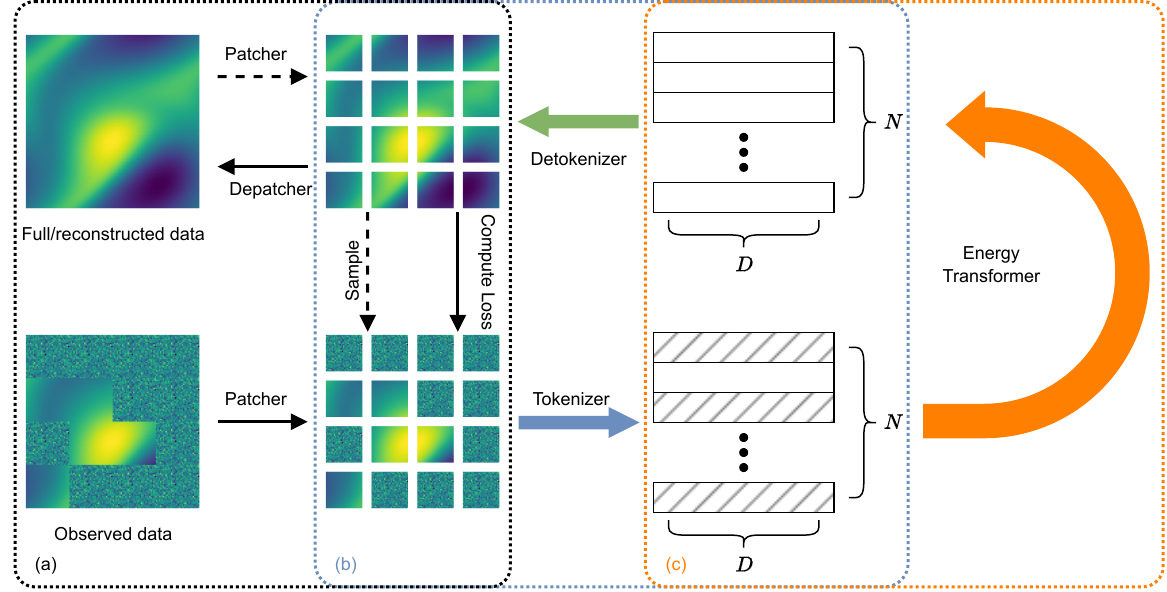}
    \caption{Overview of the proposed workflow. The dashed arrows represent the operations that are only used in the training stage. (a) Patch and depatch operations. (b) Tokenize (blue arrow) and detokenize (green arrow) operation. (c) Reconstruction operation based on the Energy Transformer.}
    \label{fig:method:workflow}
\end{figure}

The workflow of the proposed method is illustrated in Figure \ref{fig:method:workflow}.
The patcher and depatcher are fixed operations while the Energy Transformer, tokenizer and detokenizer are trainable neural networks. In the training stage, the full data are divided into patches and then tokenized, and the observed data are sampled from the full data patches and tokenized by the same tokenizer (masked patches are filled with fixed random noise and corresponding tokens are represented by dashed rectangles). Then, the Energy Transformer is trained on the tokens from the full data and observed data. In the evaluation stage, the observed data are patched and passed through the trained tokenizer, Energy Transformer and detokenizer sequentially, and then depatched to get the reconstructed data. The dashed arrows represent the operations that are only used in the training stage. In Figure \ref{fig:method:workflow}(a), we include the patch and depatch operations. Given a $d$-dimensional dataset (e.g., for images $d=2$) and patch sizes $p_1, \ldots, p_d$, the patcher splits the input data of shape $[L_1,\ldots,L_d]$ into $(L_1/p_1)\times\ldots\times(L_d/p_d)$ patches. For multi-modality data, the patches are stacked for further processing. The depatcher does the reverse. In (b) we include the tokenize (blue arrow) and detokenize (green arrow) operations. For the patched data, a neural network (tokenizer) is employed on each patch to generate vectors of length $D$, hence the tokenized data has shape $[N, D]$ where $N=(L_1/p_1)\times\ldots\times(L_d/p_d)$. The detokenizer, which is also a neural network, takes the token vector of length $D$ and produces a patch of shape $[p_1,\ldots,p_d]$. In (c) we present the reconstruction operation based on the Energy Transformer. The Energy Transformer learns to fill in the tokens corresponding to the masked patches from the knowledge of the tokens generated by the observed data.

\subsection{Energy Transformer}
\begin{figure}[H]
    \centering
    \begin{subfigure}{0.35\linewidth}
        \centering
        \includegraphics[width=\linewidth]{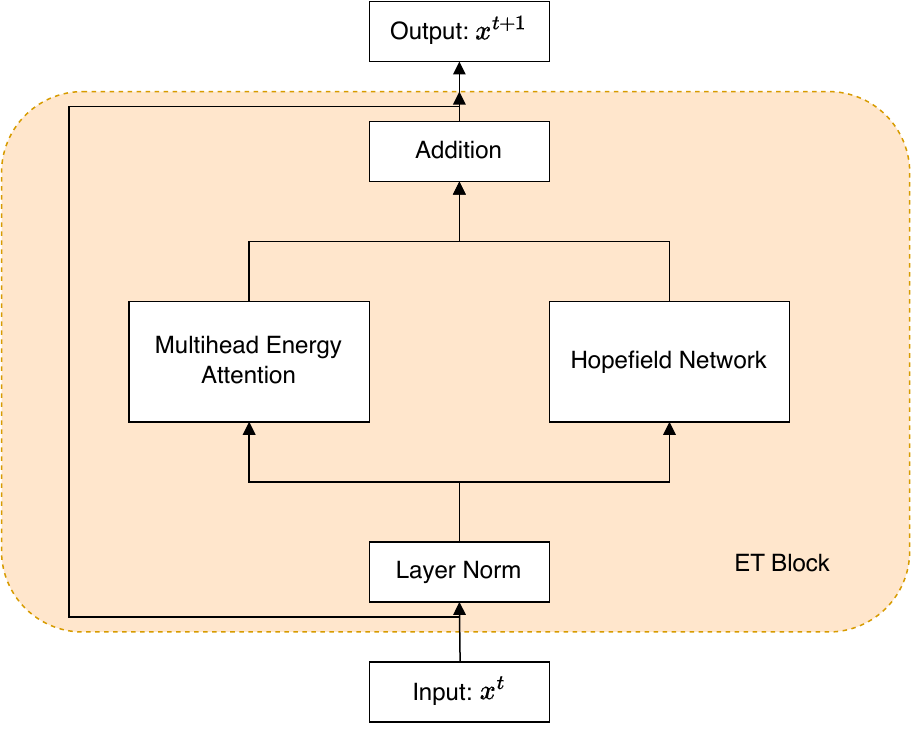}
        \caption{ET block}
    \end{subfigure}
    \hspace{0.05\linewidth}
    \begin{subfigure}{0.35\linewidth}
        \centering
        \includegraphics[width=\linewidth]{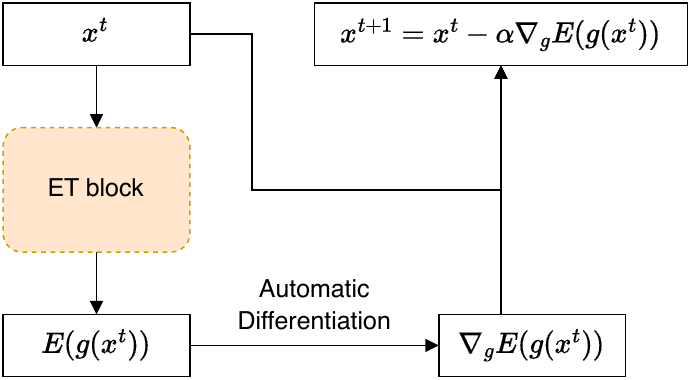}
        \caption{Update rule for ET block}
    \end{subfigure}
    \caption{ET block. (a) The structure of the ET block. (b) The update rule in the forward pass of the ET block.}
    \label{fig:method:et-block}
\end{figure}

We now introduce the structure of the Energy Transformer and how it is employed \cite{hoover2024energy}. The Energy Transformer is built on the ET block (depicted in Figure \ref{fig:method:et-block}), which consists of three components: layer normalization, energy attention layer, and Hopfield network. The energy attention layer can learn large-scale global patterns from the interaction between the input tokens, while the Hopfield network can learn small-scale local patterns within each token.

\paragraph{Layer Normalization.} Each token $x\in\mathbb{R}^D$ passed to Energy Transformer is normalized by layer normalization. This enhances the model stability and generalization ability. The layer normalization is defined as:
\begin{equation}
    g = \gamma\frac{x-\mu}{\sqrt{\sigma^2+\epsilon}}+\delta,
\end{equation}
where $\mu$ and $\sigma$ are the mean and standard deviation of the input $x$, $\gamma$ and $\delta$ are learnable parameters, and $\epsilon$ is a small constant to avoid division by zero. Furthermore, the layer normalization can be defined as the partial derivative of a Lagrangian function \cite{krotovlarge,krotov2021hierarchical}, which defines the energy function and the dynamical equations of temporal evolution leading to the decrease of the energy.

\paragraph{Multi-head Energy Attention.} The main contribution of the Energy Transformer is the multi-head energy attention layer. Similar to the conventional attention mechanism, energy attention layer learns the interaction between tokens. However, in this case, each token generates query and key vectors, and there is no value vector. The output of the energy attention layer is a scalar energy value, where the minimization of this energy indicates that the queries of the masked tokens are aligned with the keys of the observed tokens. This alignment results in the output of the energy attention layer capturing the global structure of the full data with limited observations. The energy function of the multi-head energy attention layer is defined as:
\begin{equation}
    E^{ATT} = -\frac{1}{\beta} \sum_{h=1}^{H} \sum_{C=1}^{N} \log(\sum_{B\not=C}\exp(\beta A_{hBC})),
\end{equation}
where $H$ is the number of heads, $N$ is the number of tokens, $\beta$ is a constant, and $A_{hBC}$ is the attention matrix between the $B$-th and $C$-th tokens in the $h$-th head. More specifically, the attention matrix is defined as:
\begin{equation}
    \begin{aligned}
         & A_{hBC} = \sum_{\alpha=1}^{Y} K_{\alpha hB}Q_{\alpha hC} \quad A\in\mathbb{R}^{H\times N\times N}, \\
         & K_{\alpha hB} = \sum_{l=1}^D W^K_{\alpha hl}g_{Bl} \quad K\in\mathbb{R}^{Y\times H\times N},       \\
         & Q_{\alpha hC} = \sum_{l=1}^D W^Q_{\alpha hl}g_{Cl} \quad Q\in\mathbb{R}^{Y\times H\times N},
    \end{aligned}
\end{equation}
where $W^K, W^Q\in\mathbb{R}^{Y\times H\times D}$ are learnable parameters, $g_B$ and $g_C$ are the normalized vectors of the $B$-th and $C$-th tokens, $D$ is the dimension of each token, and $Y$ is the internal dimension of the $K$-$Q$ contraction.

\paragraph{Hopfield Network.} The Hopfield network is employed to learn the local patterns within each token, ensuring that the token representations are consistent with real data. The energy function of the Hopfield network is defined as:
\begin{equation}
    E^{HN} = -\sum_{B=1}^N\sum_{\mu=1}^K G\Big(\sum_{j=1}^D \xi_{\mu j}g_{Bj}\Big),
\end{equation}
where $K$ is the number of hidden units in the Hopfield network, $\xi\in\mathbb{R}^{K\times D}$ is a learnable matrix, which corresponds to ``memories'' in simple Hopfield networks, and $G(\cdot)$ is the integral of the activation function. The Hopfield network functions analogously to the feedforward MLP in traditional transformer models, and it is applied to each token individually.

\paragraph{Forward Pass.} The patches are initially tokenized by an embedding layer. The input tokens $x$ are then normalized by a layer normalization layer. Both the multi-head energy attention layer and the Hopfield network take the normalized  $g$ and output a scalar energy. The energy of the ET block, denoted as $E$, is the summation of these two energies. Then, the output of the ET block is the local minimizer of the energy function $E$, which is obtained by iteratively updating the tokens according to the rule: $x^{t+1} = x^{t} - \alpha\nabla_g E(g(x^{t}))$, where $\alpha$ is the step size constant. The tokens are initialized as $x^{0} = x$ and iterated until convergence. In practice, the convergence is achieved after a certain number of iterations. Finally, the tokens are detokenized to obtain the reconstructed data.

\section{Results}
In this section, we demonstrate the effectiveness of the proposed method for three different examples in fluid mechanics. These examples validate the ability of the Energy Transformer to learn the reconstruction operator for different types of data with small errors.
In all these examples, two-layer MLPs with ReLU activation are used for the tokenizer and detokenizer. The patches are flattened to be tokenized. The hyperparameters for the model are in Table \ref{tab:experiments:hyperparameters}.
\begin{table}[H]
    \centering
    \begin{tabular}{ccc}
        \toprule
        Type                                & Hyperparameter                  & Value      \\
        \midrule
        \multirow{4}{*}{Patch and tokenize} & Patch size                      & $8\times8$ \\
                                            & Token dim $D$                   & 256        \\
                                            & Tokenizer hidden dim            & 256        \\
                                            & Detokenizer hidden dim          & 256        \\
        \midrule
        \multirow{3}{*}{Energy Transformer} & Heads of energy attention layer & 8          \\
                                            & Hopfield type                   & softmax    \\
                                            & Hopfield hidden dim             & 2048       \\
        \midrule
        \multirow{2}{*}{Update rule}        & Iteration steps $T$             & 12         \\
                                            & $\alpha$                        & 0.1        \\
        \midrule
        \multirow{2}{*}{Training}           & Optimizer                       & Adam       \\
                                            & Batch size                      & 16         \\
        \bottomrule
    \end{tabular}
    \caption{Model hyperparameters (the same for all three examples).}
    \label{tab:experiments:hyperparameters}
\end{table}

\subsection{2D Vortex Street}
\begin{figure}[H]
    \centering
    \includegraphics[width=0.5\linewidth]{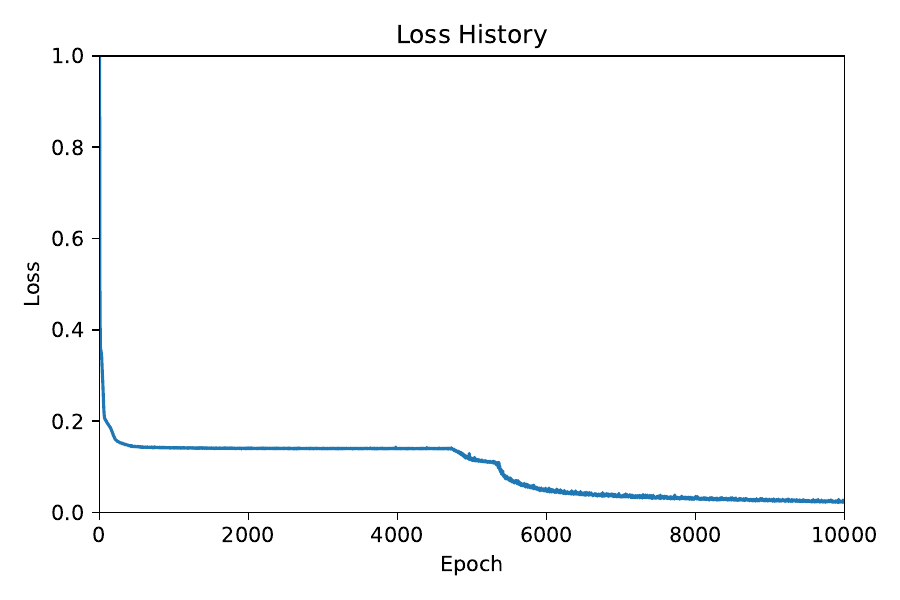}
    \caption{Training loss of the 2D vortex street example.}\label{fig:experiments:2d-vortex-street:loss}
\end{figure}

In flow past bluff bodies the von Karman vortex street is formed in the wake above a certain Reynolds number, e.g. $Re\approx 50$ for flow past a circular cylinder  \cite{raissi2019vortex}. Here, we will use simulated data to demonstrate the ET-based flow field reconstruction.
The spatial domain is $[-4,12]\times[-4,4]$, with the cylinder diameter $D=1$ and the Reynolds number is 400. The flow is simulated by a high-order spectral element method and we obtain 100 snapshots\cite{raissi2019vortex}. The training data is the first 80 time steps and the model is tested on the rest 20 time steps. The data shape of each snapshot is $[128, 256, 4]$, which is the stack of the velocity field, pressure field and the temperature field. Each patch is of shape $[8,8,4]$ due to the multi-modality of the data. The observed data is randomly masked with 90\% of the patches, in other words, only 10\% of the patches are observed. The model is trained based on the  relative mean squared error (RMSE) loss for 10,000 epochs with learning rate $10^{-4}$. The training loss history is depicted in Figure \ref{fig:experiments:2d-vortex-street:loss}, which shows that the model converges well. On a single NVIDIA A6000 GPU, the training time is around 90 minutes and the inference time is around 0.0027 seconds per snapshot.

\begin{figure}[H]
    \centering
    \includegraphics[width=0.75\linewidth]{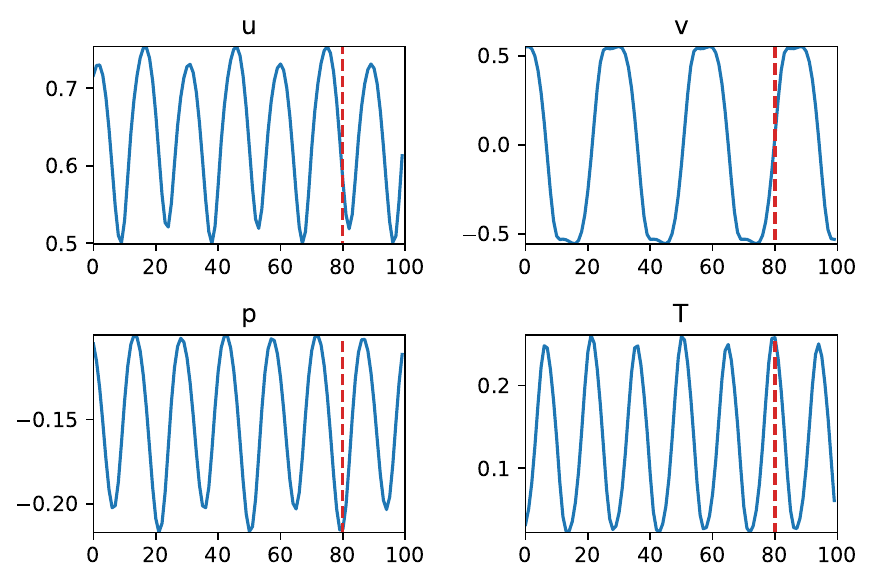}
    \caption{Signature of the 2D vortex street dataset: values of $u, v, p, T$ (horizontal velocity, vertical velocity, pressure and temperature) at the center of the domain versus the time. The red dashed line represents the split of training and test data. Periodicity can be observed in all components, which is characteristic of the vortex street.}\label{fig:experiments:2d-vortex-street:signature}
\end{figure}

\begin{figure}[H]
    \centering
    \begin{subfigure}{0.85\linewidth}
        \centering
        \includegraphics[width=\linewidth]{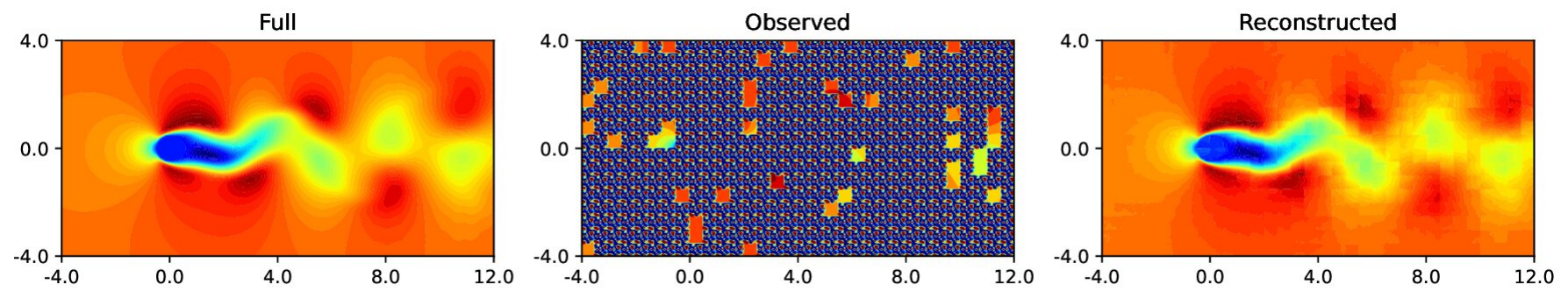}
        \caption{Velocity component $u$ (velocity along horizontal spatial coordinate).}
    \end{subfigure}
    \begin{subfigure}{0.85\linewidth}
        \centering
        \includegraphics[width=\linewidth]{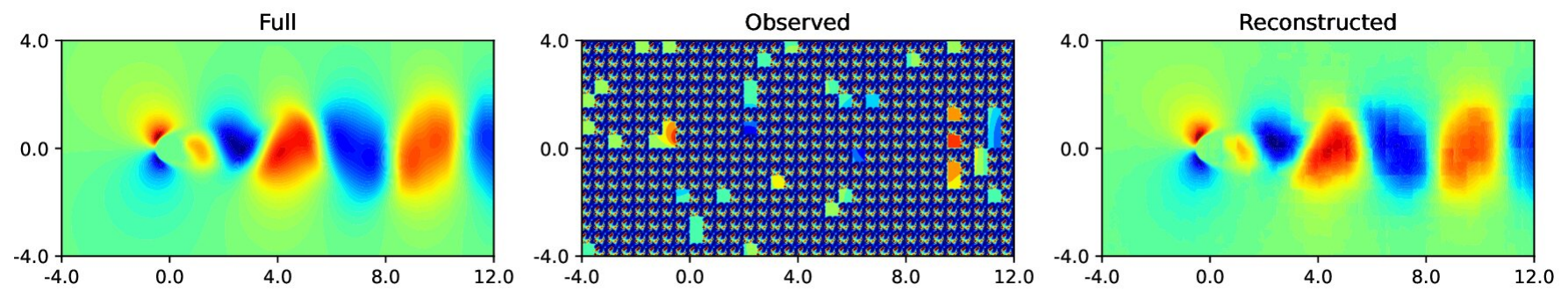}
        \caption{Velocity component $v$ (velocity along vertical spatial coordinate).}
    \end{subfigure}
    \begin{subfigure}{0.85\linewidth}
        \centering
        \includegraphics[width=\linewidth]{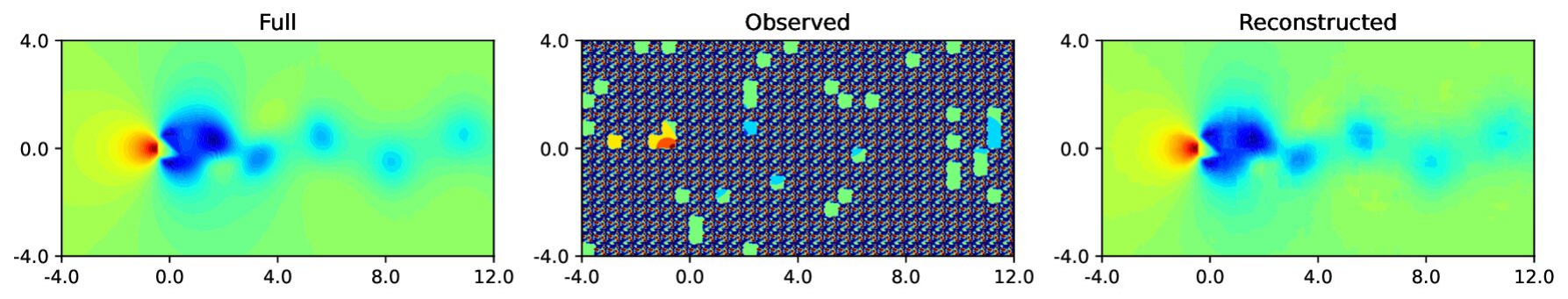}
        \caption{Pressure field $p$.}
    \end{subfigure}
    \begin{subfigure}{0.85\linewidth}
        \centering
        \includegraphics[width=\linewidth]{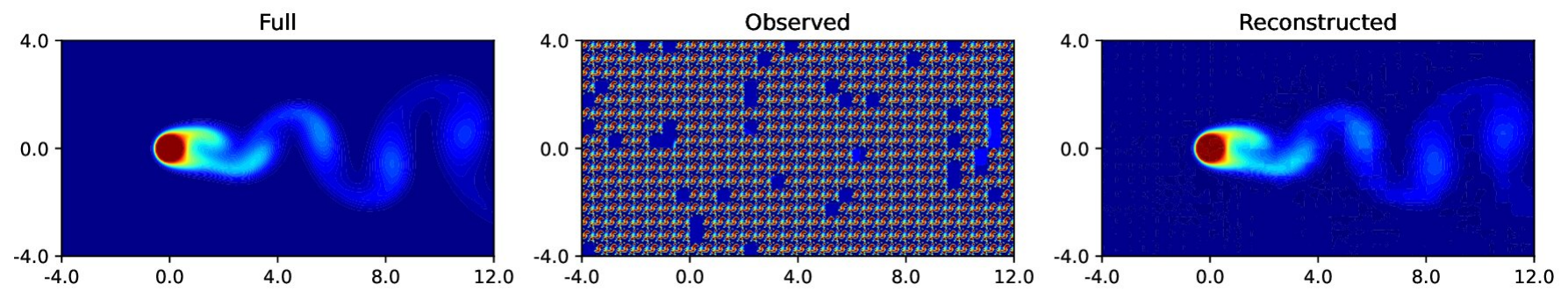}
        \caption{Temperature field T.}
    \end{subfigure}
    \caption{
        Reconstruction results of a 2D vortex street flow at time step 90. Each row in the figure displays a different physical field of the flow. In each panel, the x-axis represents the horizontal spatial coordinate, while the y-axis represents the vertical spatial coordinate, with values in physical units relevant to the flow domain. The columns show (from left to right): (1) the full field, representing the original, high-resolution simulation data; (2) the observed field, displaying sparse measurements of the field, typically available from limited sensors or measurements; and (3) the reconstructed field, where the sparse observations have been processed by an algorithm to estimate the full field. The color in each plot represents the magnitude of each physical field, with warmer colors (e.g., red) indicating higher values and cooler colors (e.g., blue) indicating lower values. Time step 90 is chosen because it is in the test dataset and it captures a characteristic snapshot of the vortex street pattern, where complex flow structures are fully developed, making it a representative example for illustrating the algorithm's performance in reconstructing detailed flow features.
    }\label{fig:experiments:2d-vortex-street:results}
\end{figure}

\begin{figure}[H]
    \centering
    \includegraphics[width=0.75\linewidth]{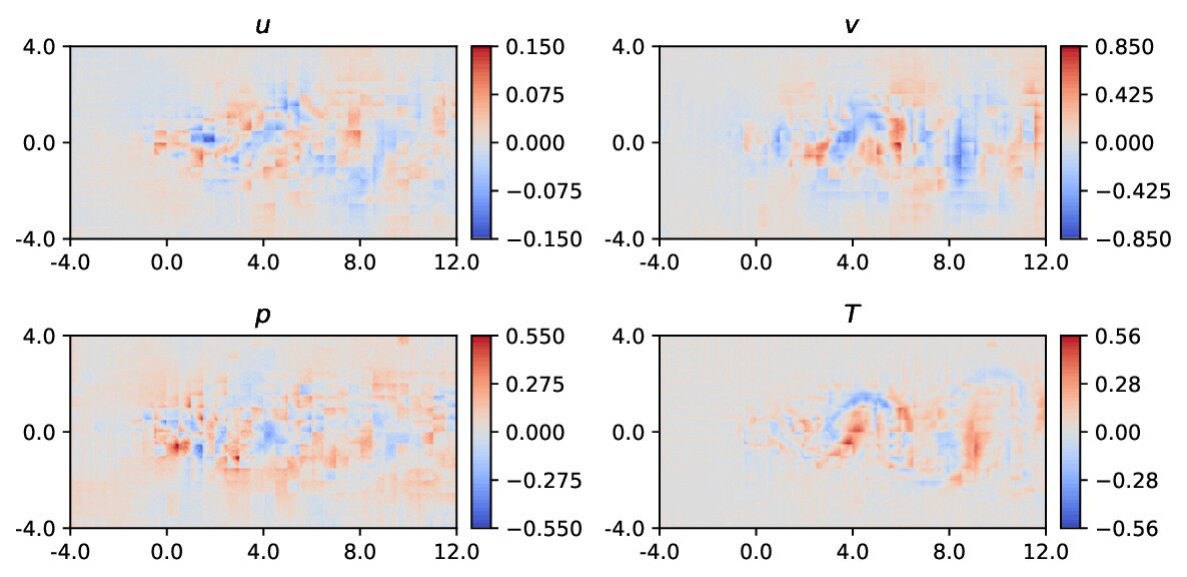}
    \caption{Errors of the 2D vortex street example at time step 90 for four physical component. The error is defined as $(\hat{x}-x)/\norm{x}_2$, $x$ and $\hat{x}$ are the ground truth and reconstructed results. Here the x and y axis are the same as in Figure \ref{fig:experiments:2d-vortex-street:results}. The colors close to red and blue indicate large positive and negative errors respectively. Errors of the $v$ component is most significant, which is a common phenomenon in vortex reconstruction\cite{jin2021nsfnets}.}\label{fig:experiments:2d-vortex-street:errors}
\end{figure}

Reconstruction results at time step $90$ are shown in Figure \ref{fig:experiments:2d-vortex-street:results}. The errors of reconstruction are shown in Figure~\ref{fig:experiments:2d-vortex-street:errors}.


The relative mean squared error (RMSE) and the relative error (defined as the RMSE divided by the $l^2$ norm of the full data) on the test set (the last 20 snapshots) are listed in Table \ref{tab:experiments:2d-vortex-street:errors}. The errors are small, which indicates the reconstruction operator is learned effectively in this example.

\begin{table}[H]
    \centering
    \begin{tabular}{ccc}
        \toprule
                  & RMSE   & Relative error \\
        \midrule
        $\vec{u}$ & 0.0286 & 0.0405         \\
        $p$       & 0.0081 & 0.0798         \\
        $T$       & 0.0080 & 0.0667         \\
        \bottomrule
    \end{tabular}
    \caption{Reconstruction errors of the 2D vortex street example. $\vec{u}=(u,v)$.}
    \label{tab:experiments:2d-vortex-street:errors}
\end{table}

\paragraph{Ablation Study.}
To evaluate the data efficiency of our method, we train the model using reduced subsets of the training data, specifically the first 25\%, 50\%, and 75\% of the full dataset. The results are presented in Table \ref{tab:experiments:2d-vortex-street:ablation}. The relative reconstruction error remains low even when only the first 50\% of the training data is used. However, the error increases significantly when the training data is further reduced to the first 25\% of the original dataset. Since the first 25\% of the data does not contain a complete period (as illustrated in Figure \ref{fig:experiments:2d-vortex-street:signature}), it is not surprising that the model fails to capture the periodicity, as it lacks sufficient information from the limited data.

\begin{table}[H]
    \centering
    \begin{tabular}{ccccc}
        \toprule
                  & 100\%  & 75\%   & 50\%   & 25\%   \\
        \midrule
        $\vec{u}$ & 0.0405 & 0.0494 & 0.0595 & 0.1616 \\
        $p$       & 0.0798 & 0.1220 & 0.1526 & 0.3116 \\
        $T$       & 0.0667 & 0.0985 & 0.1218 & 0.2493 \\
        \bottomrule
    \end{tabular}
    \caption{Relative reconstruction errors with fewer training data.}\label{tab:experiments:2d-vortex-street:ablation}
\end{table}

\subsection{2D Jet Flow}
\begin{figure}[H]
    \centering
    \includegraphics[width=0.8\linewidth]{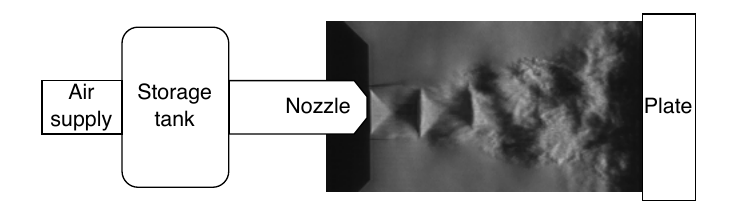}
    \caption{Illustration of the supersonic jet impinging experiment setup.}\label{fig:experiments:2d-jet:setup}
\end{figure}

The jet impinging experiments provide detailed data about the interaction between fluid and surface. In this example, the dataset is captured using Schlieren imaging techniques, hence represents the gradient of the density of the fluid~\cite{settles2017schlieren}. The experiments are conducted in the following way: the air is compressed and the temperature is approximately equal to ambient temperature, then air jet is utilized to impact the plate. At the same time, a conventional Z-type Schlieren system is adopted to visualize above under-expanded impinging jet flow. The Mach number of the flow at the nozzle exit is 1. The experiment setup is depicted in Figure \ref{fig:experiments:2d-jet:setup}. Further details can be found in \cite{wang2022deep}.

\begin{figure}[H]
    \centering
    \includegraphics[width=0.5\linewidth]{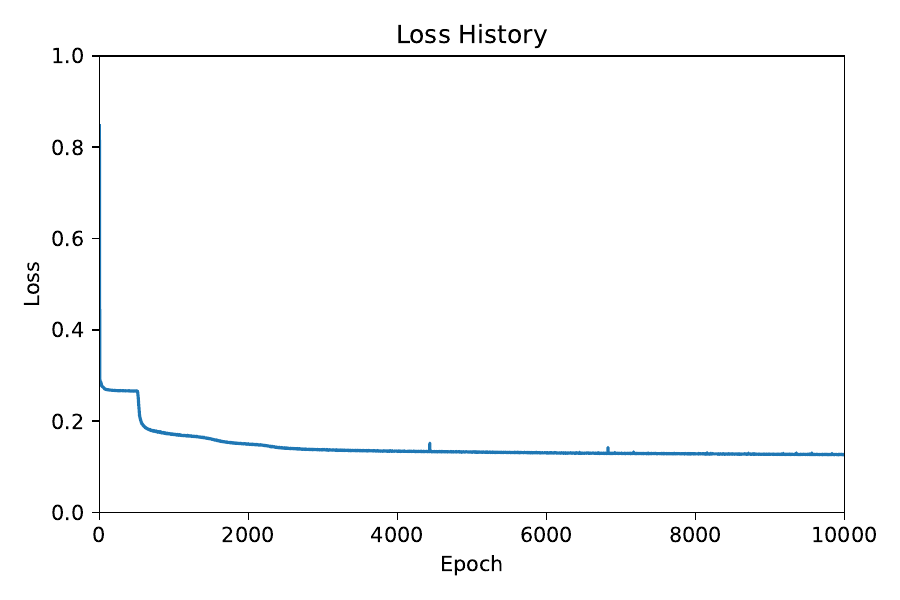}
    \caption{Training loss of the 2D jet flow example.}\label{fig:experiments:2d-jet:loss}
\end{figure}

The dataset contains 1000 snapshots of the flow field, where each snapshot has shape $[128, 256]$, and the spatial domain is $[-0.8, 5.6]\times[-1.45, 1.75]$. The values of the dataset are in the range $[0, 255]$. The observed data is randomly masked with 90\% of the patches. The model is trained with RMSE loss for 10,000 epochs with learning rate $10^{-4}$. The training loss is shown in Figure \ref{fig:experiments:2d-jet:loss}. On a single NVIDIA A6000 GPU, the training time is around 150 minutes and the inference time is around 0.0025 second per snapshot.

\begin{figure}[H]
    \centering
    \includegraphics[width=0.6\linewidth]{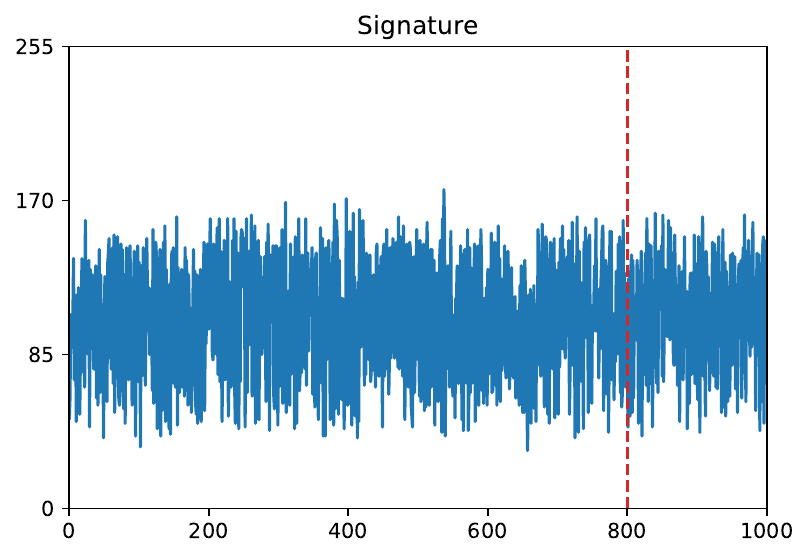}
    \caption{Signature of the 2D jet flow dataset: data values at the center of the domain versus the time. The red dashed line represents the split of training and test data.}\label{fig:experiments:2d-jet:signature}
\end{figure}

\begin{figure}[H]
    \centering
    \includegraphics[width=0.85\linewidth]{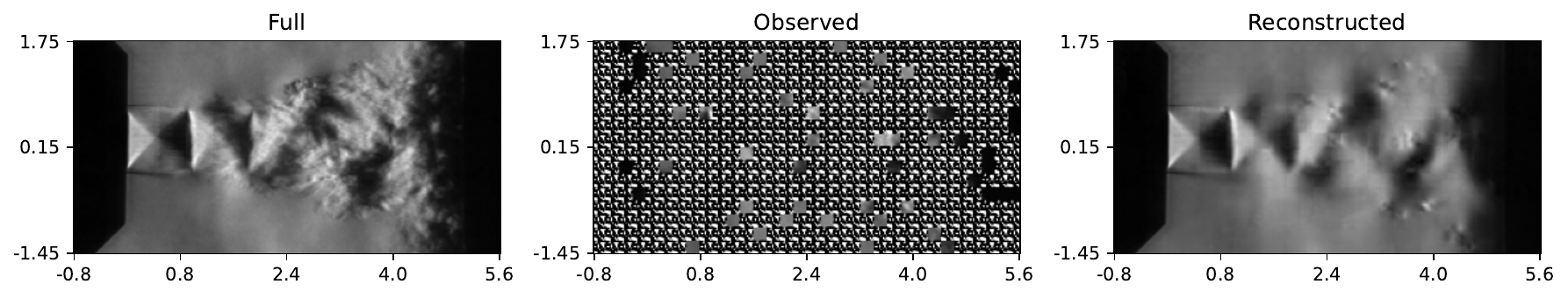}
    \caption{Results of the jet impinging example at timestep 900. The x and y axis represent the horizontal and vertical spatial coordinate. Since data has only one component, this is equivalent to image reconstruction. Time step 900 is chosen because it is in the middle of the test dataset.}\label{fig:experiments:2d-jet:results}
\end{figure}

\begin{figure}[H]
    \centering
    \includegraphics[width=0.5\linewidth]{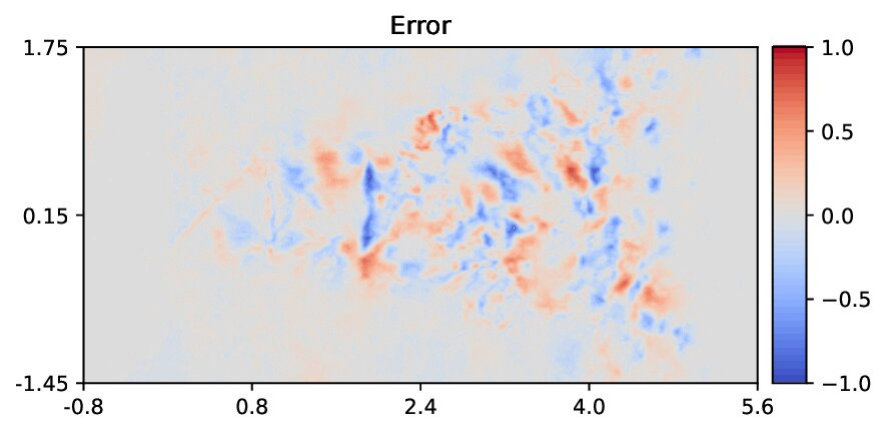}
    \caption{Errors of the jet impinging example at timestep 900. The x and y axis are the same as in Figure \ref{fig:experiments:2d-jet:results}. The errors concentraded at small-scale fluctuations.}\label{fig:experiments:2d-jet:errors}
\end{figure}

Unlike the vortex street example, the data in this example are collected from real fluid experiments, hence they contain large noise. The results and error of our methods are shown in Figure \ref{fig:experiments:2d-jet:results} and \ref{fig:experiments:2d-jet:errors}.
The RMSE is 10.4124 and the relative error is 0.1313 on the test set (last 200 snapshots). Despite the noisy nature of the data (shown in the Figure \ref{fig:experiments:2d-jet:signature}), the ET method still achieves a good performance.

\subsection{3D Turbulent Jet Flow}

Turbulent jet flow experiments facilitate the understanding of the potential core of the jet and the breakdown to turbulence downstream. In this example, the data is collected from experiments conducted at TU Delft, where the velocity measurements of jet flow can be obtained by tomographic particle tracking velocimetry (PTV). The PTV data are then processed by PINNs \cite{cai2024pinnptv} to obtain the pressure fields and continuous velocity field.

\begin{figure}[H]
    \centering
    \includegraphics[width=0.5\linewidth]{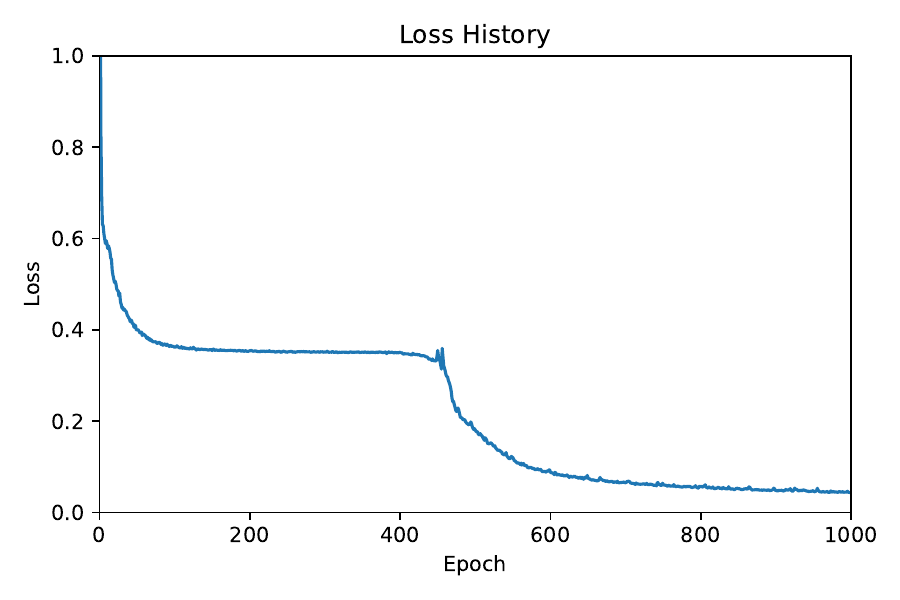}
    \caption{Training loss of the 3D turbulent jet example.}\label{fig:experiments:3d-jet:loss}
\end{figure}

\begin{figure}[H]
    \centering
    \includegraphics[width=0.75\linewidth]{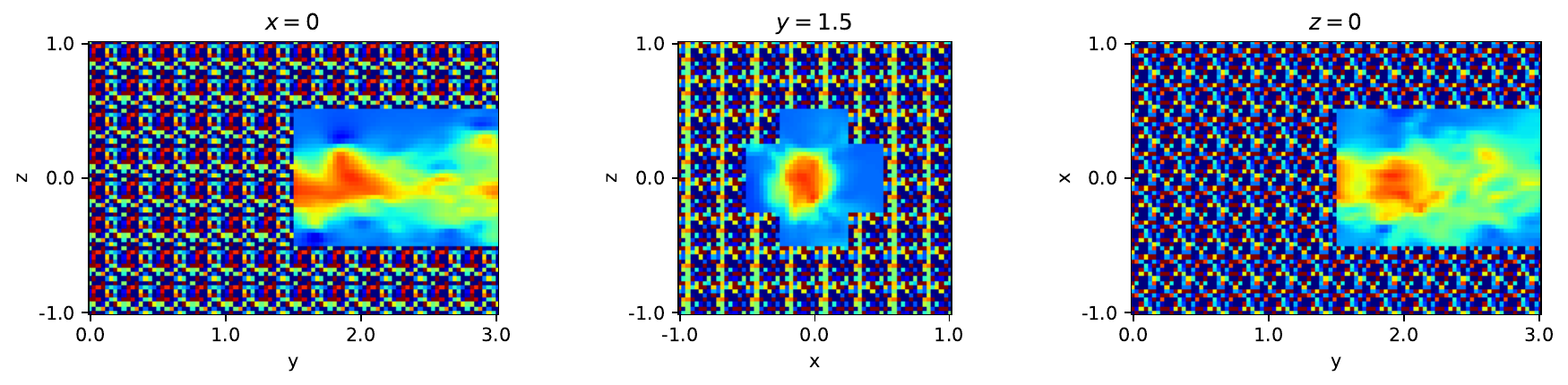}
    \caption{Illustration of the masks in the 3D turbulent jet flow example.}\label{fig:experiments:3d-jet:mask}
\end{figure}

The dataset contains 200 snapshots of the flow field, each of which has shape $[64, 96, 64, 4]$. The spatial domain is rescaled into $[-1, 1]\times[0, 3]\times[-1, 1]$. Each patch is of shape $[8, 8, 8, 4]$ due to the multi-modality of the data. The observed data is masked with a fixed pattern, where the last half of the jet flow center is observed and all the other patches are masked (illustrated in Figure \ref{fig:experiments:3d-jet:mask}). The mask rate is around 90\%. The model is trained with RMSE loss for 1,000 epochs with learning rate $10^{-3}$. The training loss is depicted in Figure \ref{fig:experiments:3d-jet:loss}. On a single NVIDIA A6000 GPU, the training time is around 70 minutes and the inference time is around 0.1 second per snapshot.

\begin{figure}[H]
    \centering
    \includegraphics[width=0.6\linewidth]{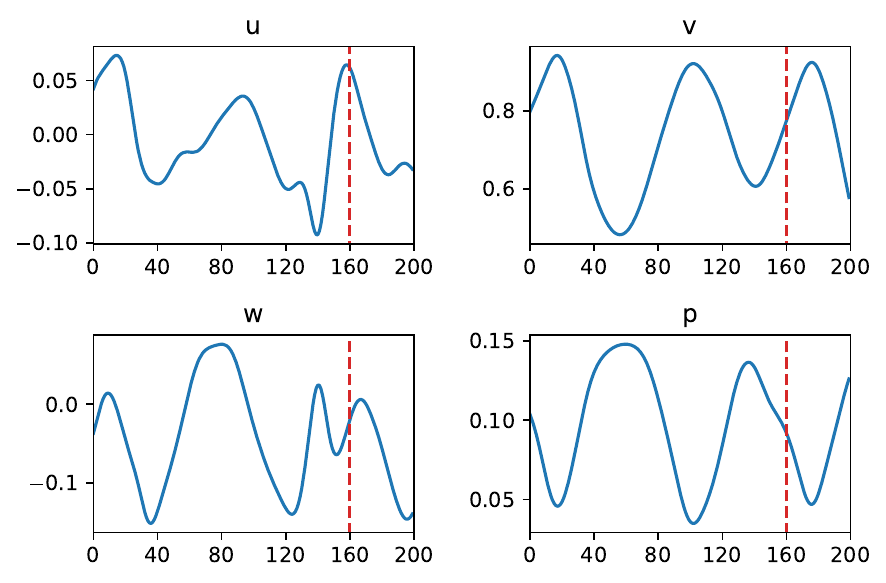}
    \caption{Signature of the 3D jet flow dataset: values of $u,v,w,p$ (velocity along $x$, $y$, $z$ axis and the pressure) at the center of the domain versus the time. The red dashed line represents the split of training and test data. Some periodicity can be observed in $v, p$ components but $u,w$ components are not periodic.}\label{fig:experiments:3d-jet:signature}
\end{figure}

\begin{figure}[H]
    \centering
    \begin{subfigure}{0.85\linewidth}
        \centering
        \includegraphics[width=\linewidth]{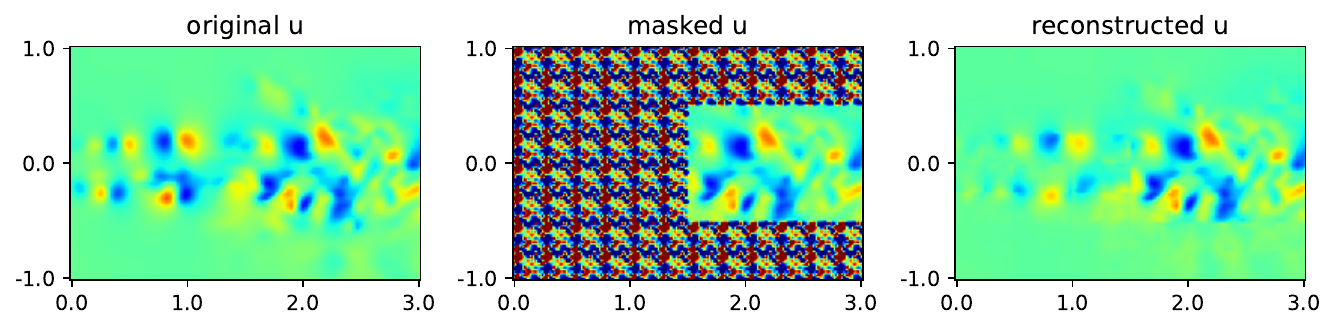}
        \caption{Velocity component $u$.}
    \end{subfigure}
    \begin{subfigure}{0.85\linewidth}
        \centering
        \includegraphics[width=\linewidth]{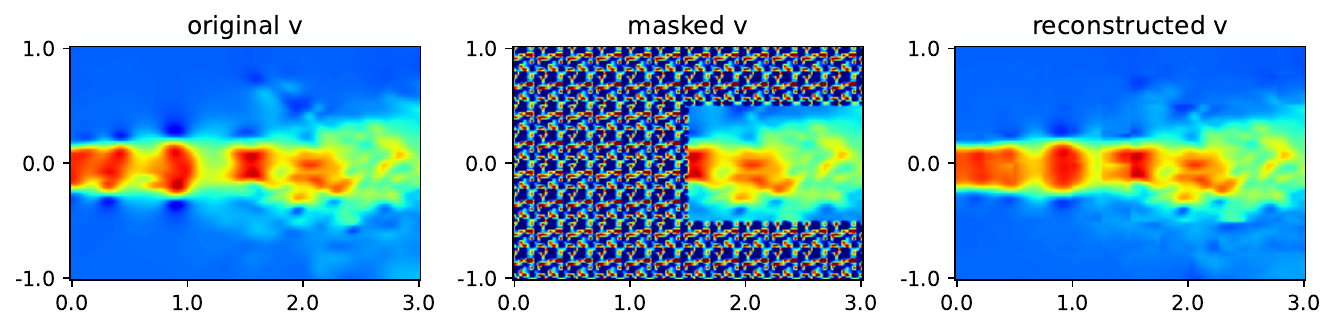}
        \caption{Velocity component $v$.}
    \end{subfigure}
    \begin{subfigure}{0.85\linewidth}
        \centering
        \includegraphics[width=\linewidth]{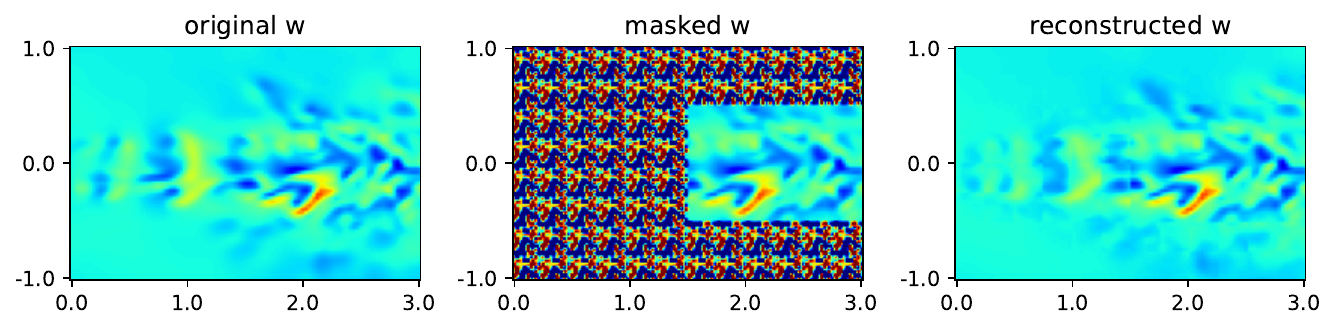}
        \caption{Velocity component $w$.}
    \end{subfigure}
    \begin{subfigure}{0.85\linewidth}
        \centering
        \includegraphics[width=\linewidth]{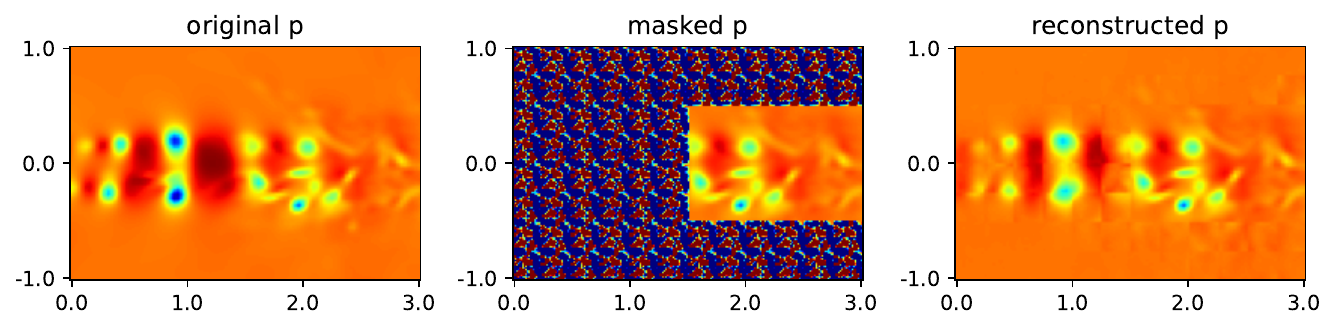}
        \caption{Pressure field $p$.}
    \end{subfigure}
    \caption{Reconstruction results of a 3D turbulent jet flow at timestep 180, visualized on the slice at $z = 0$. Each row represents a different physical field: (a) velocity component $u$ (horizontal velocity), (b) velocity component $v$ (vertical velocity), (c) velocity component $w$ (depth-wise velocity), and (d) pressure field $p$. The columns show (from left to right): (1) the full field, representing the original, high-resolution simulation data; (2) the observed field, displaying sparse measurements of the field, typically available from limited sensors or measurements; and (3) the reconstructed field, where the sparse observations have been processed by an algorithm to estimate the full field. The color in each plot represents the magnitude of each physical field, with warmer colors (e.g., red) indicating higher values and cooler colors (e.g., blue) indicating lower values. Timestep 180 is selected to illustrate a moment in the jet flow where turbulent structures are well-developed and it is in the test dataset.}\label{fig:experiments:3d-jet:results}
\end{figure}

\begin{figure}[H]
    \centering
    \includegraphics[width=0.75\linewidth]{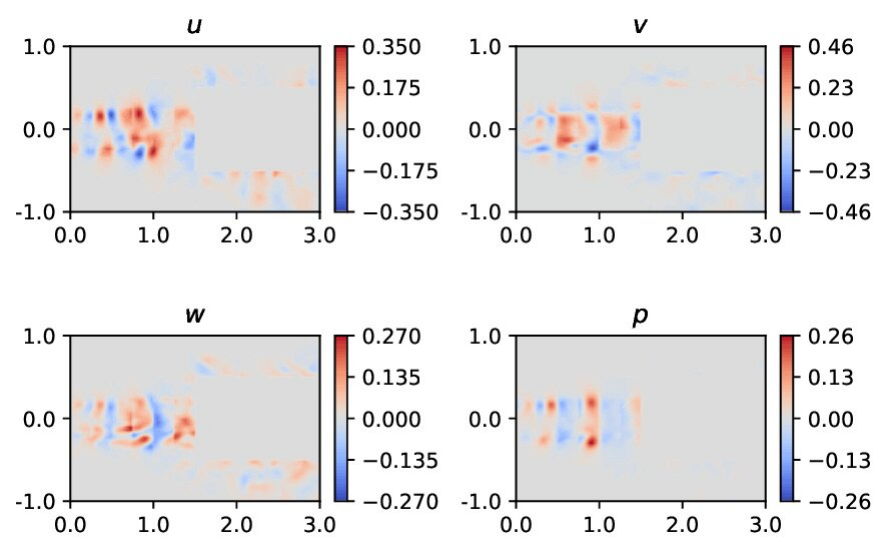}
    \caption{Errors of the 3D turbulent jet flow example at timestep 180 at the slice $z=0$. The x and y axis are the same as in Figure \ref{fig:experiments:3d-jet:results}. The errors concentrate at the upstream flow.}\label{fig:experiments:3d-jet:errors}
\end{figure}

This example is even more difficult because the dataset is in 3D and also collected from noisy experiments. The results and error of our methods are shown in Figure \ref{fig:experiments:3d-jet:results} and \ref{fig:experiments:3d-jet:errors}.

The RMSE and the relative error on the test set (the last 40 snapshots) are listed in Table \ref{tab:experiments:3d-jet}.
The relative error of $p$ is small, but the error for $\vec{u}$ is much larger. However, the relative errors are still satisfactory, because the task is essentially hard due to the 3D nature, non-periodicity (shown in Figure \ref{fig:experiments:3d-jet:signature}) and the large experimental noise.

\begin{table}[H]
    \centering
    \begin{tabular}{ccc}
        \toprule
                  & RMSE   & Relative error \\
        \midrule
        $\vec{u}$ & 0.0237 & 0.2749         \\
        $p$       & 0.0074 & 0.0993         \\
        \bottomrule
    \end{tabular}
    \caption{Reconstruction errors of the 3D turbulent jet flow example.$\vec{u}=(u,v,w)$.}
    \label{tab:experiments:3d-jet}
\end{table}



\section{Comparison with Other Methods}


The Energy Transformer presents a unique approach to reconstruction problems when compared to other operator learning methods like PINNs, DeepONet, and FNO. PINNs, which are used in Artificial Intelligent Velocimetry, can incorporate physical laws into reconstruction process. Due to their minimization nature, they can handle observed data at random positions. However, the inference cost is high as the process requires training the model to minimize the physics equation loss and data loss, which are highly non-convex. Furtheremore, PINNs suffer from poor generalization ability for extrapolation tasks.

On the other hand, DeepONet and FNO are purely data-driven methods. Their inference cost is much lower because the inference does not include minimization and is just a feedforward pass. However, they cannot handle observed data at random positions as covering all possible configurations is prohibitively expensive.

The Energy Transformer integrates the benefits of these approaches while addressing their limitations. Unlike other neural operators, Energy Transformer learns a memory-like energy function, whose minimizer is the reconstruction result. The inference is a minimization process on a relatively simple function, which is more computationally efficient than PINNs. And due to the strong inductive bias of the simple form of energy function, it can handle random observed data positions effectively without exhaustive search over all possible configurations. This unique capability makes the Energy Transformer a powerful tool for reconstruction tasks, offering low computational costs with flexibility for irregular sparse data measurements.

\section{Summary}
In this paper, we have proposed an operator learning framework with the Energy Transformer for reconstruction problems in fluid mechanics, where the goal is to recover the full flow field from sparse observed data. We demonstrated the effectiveness of the proposed method by three examples in fluid mechanics. The results show that the Energy Transformer can learn the reconstruction operator for different types of data with tolerable errors, even for the real problem with noisy data. Furthermore, its training and inference costs are relatively low compared with PINNs and the artificial intelligent velocimetry (AIV) method. In practice, sparse flow measurements can be accomplished in different parts of the domain, e.g. with small windows of particle image velocimetry (PIV), to cover the spatially varying dynamics across the domain. It is also possible to use direct numerical simulation for the training stage and build foundation models based on the Energy Transformer framework we presented herein, then the observed data will be obtained using selective experiments.

As a memory model, the Energy Transformer performs well when the test data resembles the training data, a scenario often encountered in fluid mechanics. However, for general evolution problems, the Energy Transformer may not perform as effectively, as it learns to complete individual snapshots rather than understand the underlying evolution rules. But even in this case, since the energy attention mechanism can learn large-scale structures in flow field, the Energy Transformer can still provide an averaged representation of the training data, which can be useful for limited extrapolation. Our  work provides a new perspective for solving reconstruction problems in fluid mechanics and offers a promising direction for other engineering and scientific applications, e.g., in solid mechanics using digital impage correlation (DIC), in geophysics, in weather prediction, and more.

\section{Acknowledgements}
The authors from Brown University acknowledge support of the DARPA-ABAQuS program.

\bibliographystyle{unsrt}
\bibliography{references}
\end{document}